\documentclass{aa}

\usepackage{graphicx}
\usepackage{xcolor}
\usepackage[position=top]{subfig}
\usepackage[fleqn]{amsmath}
\usepackage[utf8]{inputenc}
\usepackage{epsfig,amsmath,amssymb}
\usepackage[varg]{txfonts}
\usepackage{multirow}

\usepackage{tcolorbox}

\definecolor{myblue}{rgb}{0.00, 0.0, 0.9}
\definecolor{myred}{rgb}{0.90, 0.0, 0.0}
\definecolor{mygreen}{rgb}{0.0, 0.7, 0.0}
\usepackage[breaklinks,  citecolor=myblue, linkcolor=myred, urlcolor=purple, colorlinks=true, debug, baseurl=' ']{hyperref}

\titlerunning{Intermittent QPO of MAXI J1820$+$070}

\authorrunning{P. Zhang et al.}

\begin{document}

%

%
\title{Intermittent QPO properties of MAXI J1820$+$070 revealed by \emph{Insight}-HXMT}

\author{
P. Zhang\inst{1,2}\fnmsep\thanks{zhangpeng@ctgu.edu.cn}\fnmsep, R. Soria\inst{3,4,5}\fnmsep\thanks{rsoria@nao.cas.cn}\fnmsep, S. Zhang\inst{6}\fnmsep\thanks{szhang@ihep.ac.cn}\fnmsep, L. Ji\inst{7}, L. D. Kong\inst{8}, Y. P. Chen\inst{6}, S. N. Zhang\inst{6,9}, Z. Chang\inst{6},\\ M. Y. Ge\inst{6}, J. Li\inst{10,11}, G. C. Liu\inst{1,2}, Q. Z. Liu\inst{10,12}, X. Ma\inst{6}, J. Q. Peng\inst{6,9}, J. L. Qu\inst{6,9}, Q. C. Shui\inst{6,9}, L. Tao\inst{6},\\ H. J. Tian\inst{1,2,13}, P. J. Wang\inst{6,9}, J. Z. Yan\inst{12}, X. Y. Zeng\inst{1,2}
}
\institute{
College of Science, China Three Gorges University, Yichang 443002, China
\and
Center for Astronomy and Space Sciences, China Three Gorges University, Yichang 443002, China
\and
College of Astronomy and Space Sciences, University of the Chinese Academy of Sciences, Beijing 100049, China
\and
INAF-Osservatorio Astrofisico di Torino, Strada Osservatorio 20, I-10025 Pino Torinese, Italy
\and
Sydney Institute for Astronomy, School of Physics A28, The University of Sydney, Sydney, NSW 2006, Australia
\and
Key Laboratory of Particle Astrophysics, Institute of High Energy Physics, Chinese Academy of Sciences, 19B Yuquan Road, Beijing 100049, China
\and
School of Physics and Astronomy, Sun Yat-Sen University, Zhuhai 519082, China
\and
Institut f{\"u}r Astronomie und Astrophysik, Kepler Center for Astro and Particle Physics, Eberhard Karls, Universit{\"a}t, Sand 1, D-72076 T{\"u}bingen, Germany
\and
University of Chinese Academy of Sciences, Chinese Academy of Sciences, Beijing 100049, China
\and
CAS Key Laboratory for Research in Galaxies and Cosmology, Department of Astronomy, University of Science and Technology of China, Hefei 230026, China
\and
School of Astronomy and Space Science, University of Science and Technology of China, Hefei 230026, China
\and
Key Laboratory of Dark Matter and Space Astronomy, Purple Mountain Observatory, Chinese Academy of Sciences, Nanjing 210008, China
\and
School of Sciences, Hangzhou Dianzi University, Hangzhou 310018, China
}
          
\date{Received XXX / Accepted XXX}

\abstract{
We investigate the dynamical properties of low frequency quasi-periodic oscillations (QPOs) observed from the black hole X-ray binary MAXI J1820$+$070 during the early part of its 2018 outburst, when the system was in a bright hard state. To this aim, we use a series of observations from the Hard X-ray Modulation Telescope \emph{Insight}-HXMT, and apply a wavelet decomposition (weighted wavelet Z-transforms) to the X-ray light-curve. We find that the QPO phenomenon is intermittent within each individual observation, with some sub-intervals where the oscillation is strongly detected (high root-mean-square amplitude) and others where it is weak or absent. The average life time of individual QPO segments is $\approx$ 5 oscillation cycles, with a 3 $\sigma$ tail up to $\approx$ 20 cycles. There is no substantial difference between the energy spectra during intervals with strong and weak/absent QPOs. We discuss two possible reasons for the intermittent QPO strength, within the precessing jet model previously proposed for MAXI J1820$+$070. In the rigid precession model, intermittent QPOs are predicted to occur with a coherence Q $\approx$ a few when the disk alignment time-scale is only a few times the precession time-scale. Alternatively, we suggest that changes in oscillation amplitude can be caused by changes in the jet speed. We discuss a possible reason for the intermittent QPO strength, within the precessing jet model previously proposed for MAXI J1820$+$070: we suggest that changes in oscillation amplitude are caused by changes in the jet speed. We argue that a misaligned, precessing jet scenario is also consistent with other recent observational findings that suggest an oscillation of the Compton reflection component in phase with the QPOs. 
}


\maketitle

\section{Introduction}
\label{sec:intro}
Stellar-mass black holes (BHs) in low-mass X-ray binaries accrete matter from their companion via Roche lobe overflow and undergo occasional outbursts. The outburst evolution reflects the change of the balance between the thermal emission from the disk and the non-thermal emission from either a corona or a jet. During outbursts, such systems go through a series of spectral states, including the low hard state (LHS), the hard intermediate state (HIMS), the soft intermediate state (SIMS) and the high soft state (HSS) \citep{Homan2005,Belloni2010,Belloni2016}.

An important observational property of stellar-mass BH X-ray binaries, potentially a probe of the inflow structure just outside the horizon, is the presence of  quasi-periodic oscillations (QPOs) \citep{Remillard2006}. QPOs are distinct peaks in the power density spectra of their X-ray light curves. QPOs are usually classified primarily into high-frequency ($\sim$ 10 -- $10^{3}$ Hz) and low-frequency ($\sim$ 10$^{-2}$ -- 10 Hz). The latter are further divided into sub-types (Type-A, Type-B and Type-C: \citealt{Casella2004}), based on their coherence and on the strength of different frequency \citep{Wijnands1999, Homan2001, Remillard2002}. The coherence is expressed by the quality factor Q $= \nu/\Delta\nu$, where $\nu$ is the centroid frequency and $\Delta\nu$ is the full width at half maximum near the centroid frequency. The most common QPO sub-type (also the one with the highest Q factor) is the Type-C; usually, it has a frequency in the range of a few mHz to 10 Hz, but on occasions it has been detected up to 30 Hz \citep{Revnivtsev2000}. 
Explanations for the origin of Type-C QPOs are still controversials; alternative models include inner-disk instabilities and Lense-Thirring (LT) precession of either the jet or the inner hot flow/corona \citep{Stella1999, Ingram2009, Ingram2011}.

The Galactic X-ray transient MAXI J1820$+$070 is a particularly suitable target for the study of Type-C QPOs. It is a dynamically confirmed Galactic BH
\citep{Torres2019}, discovered by the Monitor of All-sky X-ray Image (MAXI) on 2018 March 11 \citep{Kawamuro2018}. The position of the transient is consistent with that of ASASSN-18ey, an optical transient discovered 5 days earlier \citep{Denisenko2018}. {\it Gaia} also detected an optical counterpart at the J2000 position of R.A.~$=$ 18$^h$20$^m$21$^s$.94, Dec.~$=$ $+$07$^{\circ}$11$^{\prime}$${07}\farcs{19}$, with an apparent brightness $g \approx 17.41$ mag. Its distance was estimated as $3.46^{+2.18}_{-1.03}$ kpc, from the {\it Gaia} Data Release 2 data \citep{Gandhi2019}. \cite{Atri2020} provided a consistent and more precise distance measurement of (2.96 $\pm$ 0.33) kpc, using the parallax obtained from radio interferometry. The mass of the BH in MAXI J1820$+$070 was originally estimated as $\approx$ 6 -- 8 $M_{\odot}$ \citep{Torres2019} from optical spectroscopic studies, and later revised to $\approx$ 8 -- 9 $M_{\odot}$ \citep{Torres2020}
based on \cite{Atri2020}'s distance and inferred jet orientation angle of $63^{\circ} \pm 3^{\circ}$. The mass donor is a low-mass star with $M_2 \approx (0.6 \pm 0.1) M_{\odot}$ \citep{Torres2020}.

MAXI J1820$+$070 remained in a hard state throughout the initial outburst phase (2018 March -- June) \citep{Shidatsu2018,Shidatsu2019}. As expected in the hard state, a large number of type-C QPOs were detected \citep{Uttley2018,Bright2018}. Detailed X-ray timing studies during this hard state outburst have been presented in several recent works \citep[{\it e.g.},][]{Paice2019,Paice2021,Dzielak2021,Axelsson2021,DeMarco2021,Mao2022,Thomas2022,Prabhakar2022,Zhou2022,Gao2023,Kawamura2023}. 

In this work, we focus on the timing information from the X-ray satellite \emph{Insight-Hard X-ray Modulation Telescope} ({\it Insight-HXMT}). With {\it Insight-HXMT}, QPOs were detected at energies as high as 250 keV during the LHS; a precessing jet scenario was proposed as a possible explanation \citep{Ma2021}. Here, we revisit the \emph{Insight}-HXMT observations during the LHS, and apply  wavelet decomposition to the light curves, to determine the dynamic properties of the QPOs. We describe the observations and data analysis in Section 2, present the main results in Section 3, and discuss possible interpretations in Section 4. 

\begin{figure}
\centering
\includegraphics[scale=0.53]{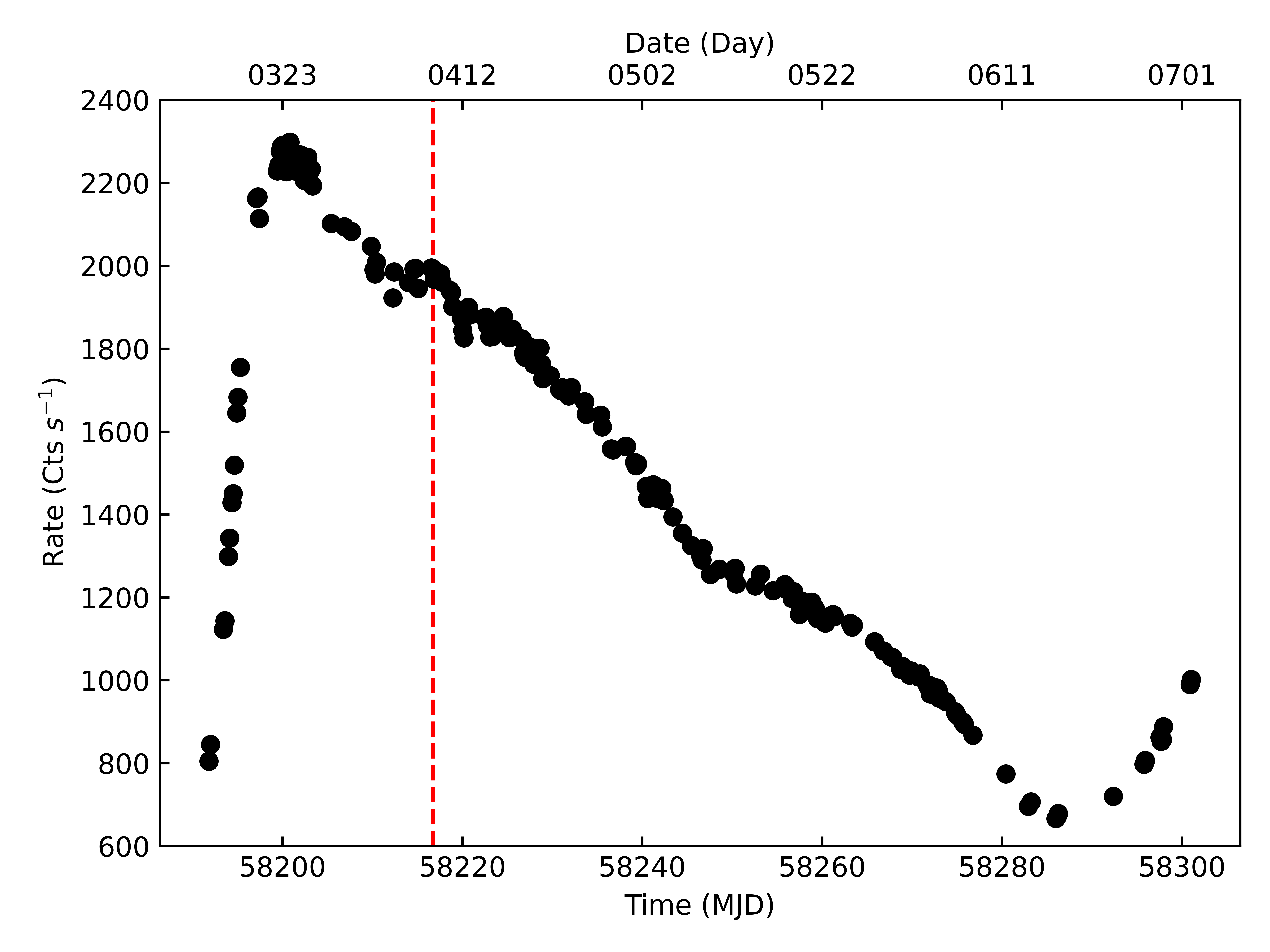}
\vspace{-0.5cm}
\caption{\emph{Insight}-HXMT/HE lightcurve of MAXI J1820$+$070 in the 30 -- 150 keV band during the 2018 LHS outburst. Calendar dates in the Month -- Day format are reported on the top horizontal axis. The dotted red line marks the time of ObsID P01146610150, chosen for the dynamic QPO study in this work.
}
\label{time_count}
\end{figure}

\section{Observations and Data Analysis}
\label{sec:Observations}

\emph{Insight}-HXMT \citep{Zhang2014,Zhang2020} was launched on 2017 June 15 and is the first Chinese X-ray astronomical satellite. It carries three X-ray instruments: the High Energy (HE; \citealt{Liu2020}), Medium Energy (ME; \citealt{Chen2020}) and Low Energy (LE; \citealt{Cao2020}) X-ray Telescopes. All three work in a collimated way and are equipped with blind detectors. The HE is made of 18 cylindrical NaI(TI)/CsI(Na) phoswich detectors, and covers the 20.0 -- 250.0 keV band, with a total detection area of 5100 cm$^{2}$. The ME consists of 1728 Si-PIN detectors, sensitive in the 5.0 -- 30.0 keV band, with a total detection area of 952 cm$^{2}$. The LE operates with swept charged devices in the 1.3 -- 15.0 keV band, and has a total detection area of 384 cm$^{2}$.

MAXI J1820$+$070 was monitored by \emph{Insight}-HXMT 146 times for a total exposure time of $2.56$ Ms. Sixty observations for a total exposure of 1.56 Ms were taken during the LHS (2018 March 14 -- July 6). We reduced the LHS data following the standard procedure in the \emph{Insight}-HXMT data analysis software {\sc HXMTDAS}, version v2.02\footnote{http://hxmt.org/index.php/usersp/dataan}. The good time intervals were selected with the following criteria: elevation angle $>$ 10$^{\circ}$; geomagnetic cutoff rigidity $>$ 8 GeV; pointing offset angle $<$ 0.04$^{\circ}$; $>$ 600 s away from the South Atlantic Anomaly (SAA). The background model was produced with standard python scripts (\emph{hebkgmap}, \emph{mebkgmap} and \emph{lebkgmap}) and subtracted off in both timing and spectral analyses. 

Traditional tools for time-variability studies include the analysis of the power spectral density (PSD), the short-time Fourier transform (STFT) of the light curve, and the Lomb-Scargle periodogram (LSP) \citep{Lomb1976, Scargle1982, Zechmeister2009}. PSDs and LSPs are best suited for the search of strictly periodic signals, but they are not designed for the analysis of signals with time-varying frequency. STFTs can handle variable frequencies, but are heavily dependent on the choice of window function as they often have significant side effects if chosen poorly \citep{Zhao2020}. Instead, weighted wavelet Z-transforms (WWZ\footnote{https://github.com/eaydin/WWZ}; \citealt{Foster1996,Torrence1998}) produce a robust map in the frequency--time domain, and are the best tool available for variability studies of QPOs in both strength and frequency. Examples of the use of WWZ analysis to investigate the transient nature of QPOs in various astrophysical systems are for example \cite{Bravo2014}, \cite{Benkhali2020}, \cite{Urquhart2022}.

For our study of MAXI J1820$+$070, we used a version of WWZ modified with a Morlet parent function \citep{Foster1996}. We produced three sets of 2-dimensional WWZ colour maps (power spectrum vs elapsed time), for the LE (2 -- 8 keV), ME (8 -- 30 keV) and HE (30 -- 150 keV) bands. By doing the WWZ analysis on the three bands independently, and then comparing and correlating the results, we can more confidently identify real structures (present in all three bands) as opposed to possible artifacts. 

We also extracted energy spectra over the same time intervals, using {\sc hxmtdas} tools \emph{hespecgen}, \emph{mespecgen} and \emph{lespecgen}. We used the task {\it grppha} within {\sc ftools} \citep{Blackburn1995} to rebin the spectral data to a minimum of 30 counts per bin, so that we could later fit the spectra with the $\chi^2$ statistics.
For our spectral analysis (Section 3.2), we used {\sc xspec} \citep[v.~12.9.0n, ][]{Arnaud1996}.  We fitted the LE, ME and HE spectra simultaneously. A systematic error of 1\% was added to account for the calibration uncertainties, and we also allowed for a free normalization constant between the three instruments. 

\begin{figure*}
\centering
\includegraphics[scale=0.5]{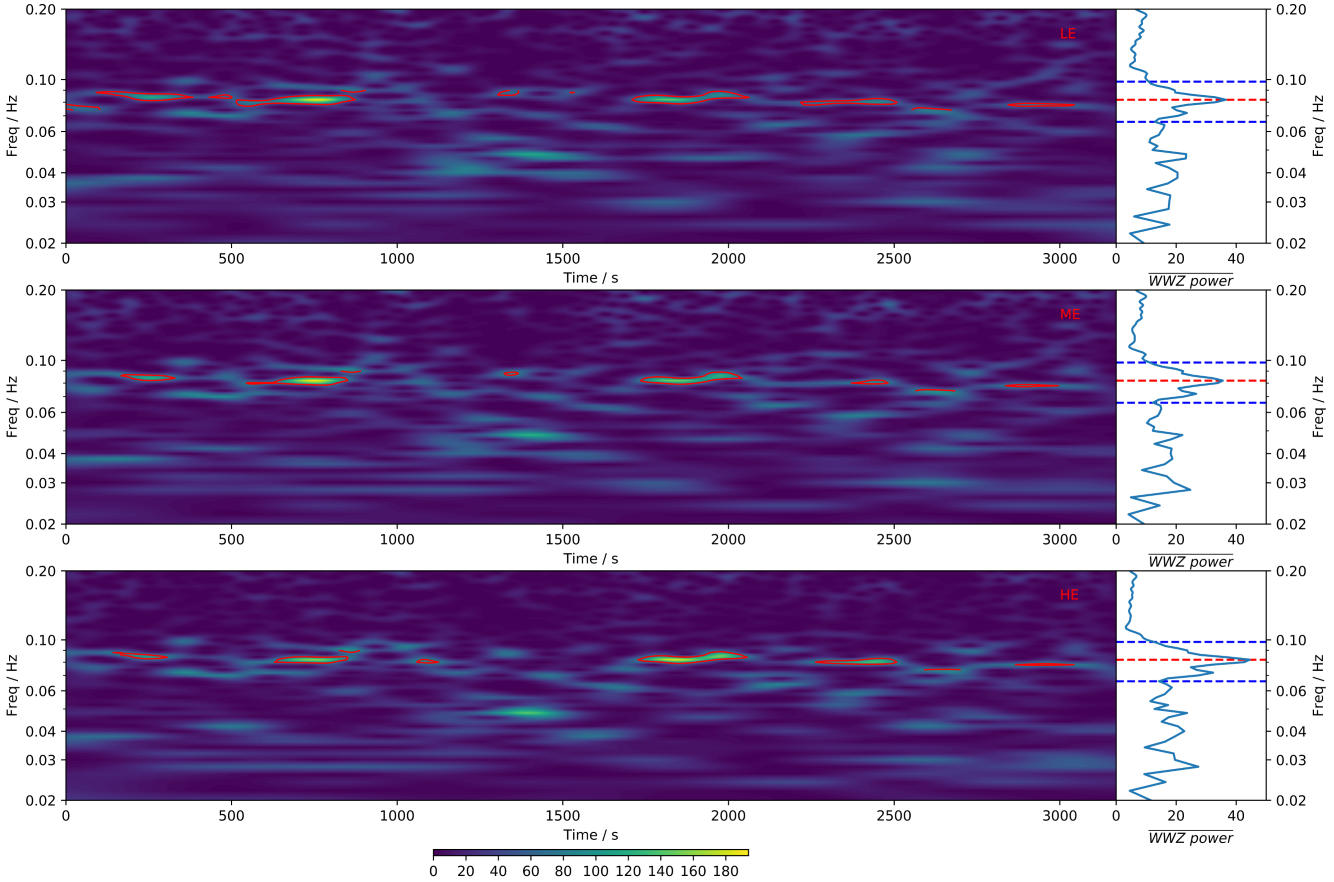}
\caption{
2-D plot on the left shows the dynamic WWZ power in time-frequency space during ObsID P011466101502. Red contours represent regions (in time-frequency space) of QPO activity with significance $>$ 3 $\sigma$. Units of WWZ power are (root-mean-square/mean)$^2$ Hz$^{-1}$. The line plot on the right-hand side of all panel shows the corresponding time-averaged WWZ power spectrum over the whole observation in each band; the dashed red line indicates the peak power at a QPO frequency of 0.082 Hz, and the dashed blue lines correspond to $\pm$ 20\% around peak frequency. 
Top panel: for the HE data in the 30 -- 250 keV band;
Middle panel: for the ME data in the 5 -- 30 keV band;
Bottom panel: for the LE data in the 1.3 -- 15.0 keV band.
}
\label{lc}
\end{figure*}

\section{Results}
\subsection{Timing results}
\label{subsec:Search qpo}
As a starting point of our time-resolved WWZ analysis, we used the observation-averaged QPO frequencies determined by \cite{Ma2021} as central frequencies for each observation. We can do that because frequencies do not vary substantially within each observation. Instead, the strength of the QPO is strongly variable within each observation: each interval of QPO detection is of short duration, and for most of the exposure time the oscillation is weak or undetected. Here, we chose ObsID P011466101502 (exposure time:3172 s, on 2018 April 08) to illustrate this general behaviour (Figure \ref{lc}).  

We determined the significance of intra-observation QPO structures by comparing their power with the distribution of power values within 20\% of the central QPO frequency. We selected all QPO intervals with a significance larger than 3 $\sigma$ (red contours in the left panels of Fig. \ref{lc}) by calculating the average and standard deviation within the region; we applied this procedure separately for the three detectors. Then, we determined the "life time" of each significant QPO interval, expressed in units of the average QPO period measured over that segment. The life time depends of course on the threshold we choose for the definition of QPO significance. If we choose a much lower significance, not only will the life time of each QPO segment increase, but neighbouring segments may connect into a single interval. However, the main qualitative finding of our analysis, that is the presence of stronger and weaker intervals of QPO behaviour, remains valid. For simplicity, we will sometimes refer to observation sub-intervals in which a QPO is or is not detected at the 3-$\sigma$ level as "QPO'' and "non-QPO" intervals.

\begin{figure*}
\centering
\includegraphics[scale=0.385]{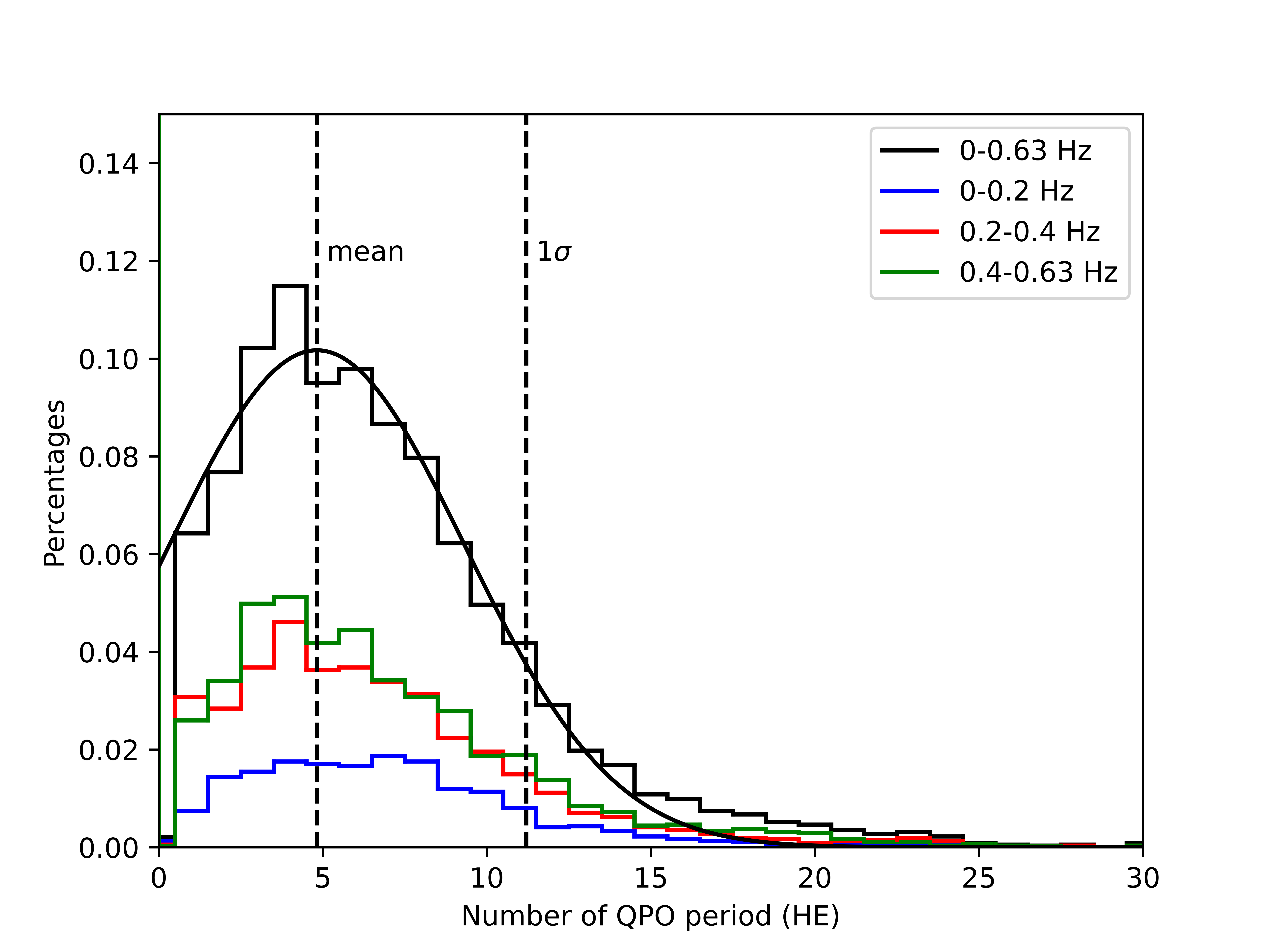}\hspace{-0.55cm}
\includegraphics[scale=0.385]{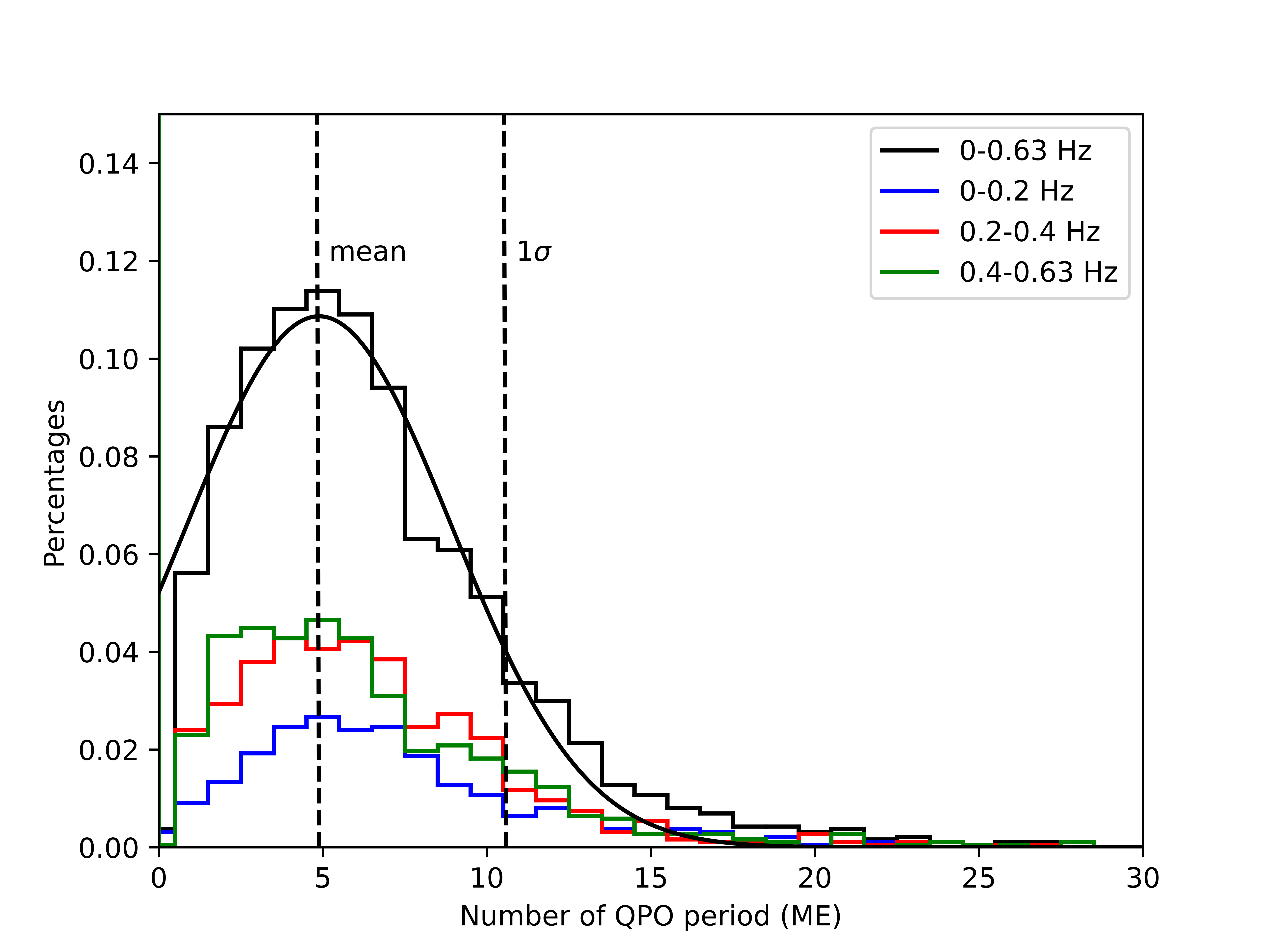}\hspace{-0.55cm}
\includegraphics[scale=0.385]{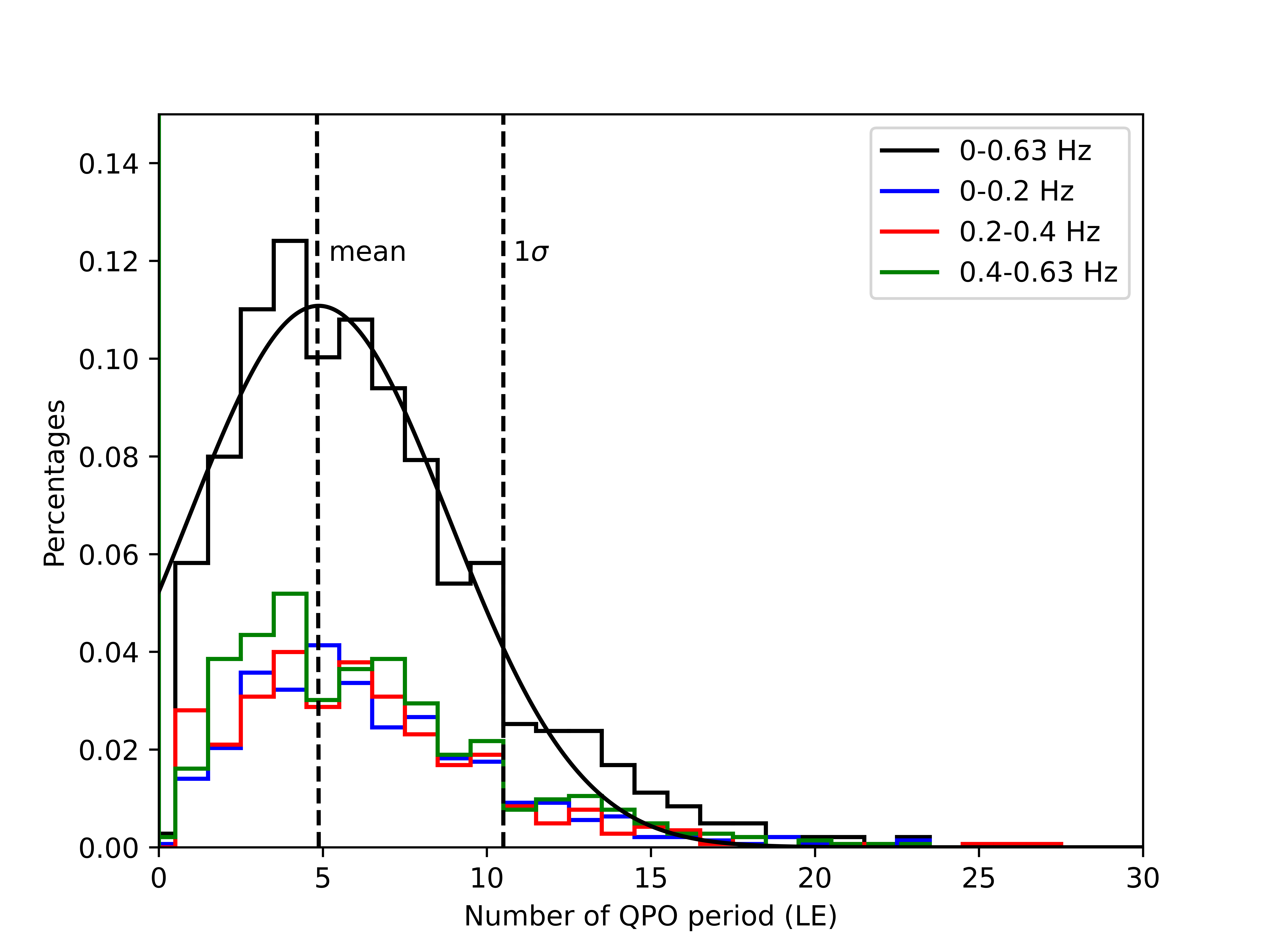}
\vspace{-0.4cm}
\caption{QPO life time distribution over all the 2018 March -- June {\it Insight-HXMT} observations, in units of cycles; the left panel is for the LE energy band, the middle panel for ME, and the right panel for HE. In all panels, the blue histogram is the life time distribution of QPOs with frequencies $<$ 0.2 Hz; the red histogram is the distribution for QPOs with frequencies of 0.2 -- 0.4 Hz; the green histogram is for QPO frequencies of 0.4 -- 0.63 Hz; the black histogram is the life time distribution for all QPOs of any frequency. The total life time distribution in each panel has been fitted with a Gaussian (solid black curve); dashed black lines mark the Gaussian mean life time and the value of 1$\sigma$ above the mean. 
}
\label{qpo1}
\end{figure*}

QPO frequencies increase with time during the LHS outburst phase \citep{Ma2021}. However, we found that the dimensionless life time of each QPO segment remains approximately constant. The average life time is 6.4 cycles in the LE band, 6.6 in the ME band, and 6.8 cycles in the HE band, and the standard deviations are 4.0, 4.4, and 4.6, respectively. Calculated over the full energy band, the average life time is $(6.7 \pm 4.4)$ cycles. If we fit the life time distributions with Gaussians (Fig.~\ref{qpo1}), the mode is (4.82 $\pm$ 0.24) cycles in the HE data, (4.89 $\pm$ 0.22) cycles in the ME data, (4.88 $\pm$ 0.26) cycles in the LE data, and (4.85 $\pm$ 0.21) over the full energy band. The standard deviations (1 $\sigma$) are 4.51, 4.03, 3.98, and 4.31 cycles, respectively. There is a slight excess of apparently long QPO segments (life time of $\approx$ 15 -- 20 cycles) over the best-fitting Gaussian distributions (Figure \ref{qpo1}); such events likely correspond to adjacent, shorter QPO intervals connected together in our WWZ analysis. Furthermore, we calculated the QPO life time distributions at different QPO frequencies (0 -- 0.2 Hz, 0.2 -- 0.4 Hz and 0.4 -- 0.63 Hz), shown as blue, red and green histograms in Figure \ref{qpo1}: the three distributions are not significantly different. We conclude that the average life time of QPO events in the LHS of MAXI J1820$+$070 is $\approx$ 5 cycles, with a standard deviation of the life time distribution of $\approx$ 4 cycles, regardless of energy band and QPO frequency. Finally, we used the {\sc ftools} \citep{Blackburn1995} task {\it powspec} to calculate the power spectral density of all observations, and the root-mean-square (rms) value of the QPOs (fitted with Lorentz functions). We used the WWZ maps to distinguish exposure time intervals with and without a QPO. We obtain an average rms $\approx$ 11.3\% for the intervals of QPO detection, and rms $<$ 9.6\% for the intervals in which no significant oscillations are detected.

To make sure that the QPO segments found in our WWZ analysis are real physical events and not just statistical artifacts, for each observation we correlated the three WWZ maps in the LE, ME and HE bands. The Pearson's correlation coefficient $\gamma_{(1,2)}$ between a band 1 and a band 2 is
\begin{equation}
    \gamma_{(1,2)} = \frac{{\mathrm{Cov}}_{(1,2)}}{\sqrt{{\mathrm{Var}}_1 {\mathrm{Var}}_2}},
	\label{eq:cc}
\end{equation}
where Var$_1$ and Var$_2$ are the variances of the WWZ powers in the two bands, and Cov$_{(1,2)}$ is their covariance. 

For each observation, we calculated the correlation coefficient over a frequency range of $\pm$20 \% of the average QPO frequency in that observation. For the specific observation P011466101502 illustrated here as an example, the frequency range for the correlation coefficient is 0.064 -- 0.096 Hz (around the average QPO frequency of 0.082 Hz). We obtain $\gamma_{LE,HE}$ = 0.84 and $\gamma_{ME,HE}$ = 0.90. This suggests that the QPO structures independently seen in the three bands at the same time (Fig.~\ref{lc}) are indeed real. In total, there are 60 \emph{Insight}-HXMT observations for which we could do a QPO correlation analysis between the three bands. The average LE-ME correlation coefficient over all those observations is $\gamma_{LE,HE} = 0.83$, with a standard deviation of their distribution $\sigma_{LE,HE} = 0.06$. For the ME-HE bands, the average correlation $\gamma_{ME,HE} = 0.90$, with a standard deviation $\sigma_{ME,HE} = 0.05$. We also carefully checked the observation epochs with relatively low Pearson's correlation coefficients, and found that their QPO signals were weak and the noise was too strong, which lowered the significance of the Pearson's correlation coefficient. We removed from our estimate the 11 observations with lowest QPO signals (significance $\le$ 3 $\sigma$), and re-calculated the band-to-band correlations over the remaining 49 observations with stronger QPO detection. This gives $\gamma_{LE,HE} = 0.90$ with a standard deviation of their distribution $\sigma_{LE,HE} = 0.04$ and $\gamma_{ME,HE} = 0.94$ with a standard deviation of their distribution $\sigma_{LE,HE} = 0.03$, respectively.

In order to check that the measured correlations are not an artifact of wavelet analysis, we performed the following test.  We assumed the same stationary power spectrum for all three bands. We used Fourier transforms to generate 1000 simulated light curves for each band. We also generated independent backgrounds and added 5\% counting noise for each light curve. Then, we applied wavelet analysis to the simulated light curves, and selected the same frequency range to calculate the correlation coefficients. We obtained correlation coefficients consistent with 0: for LE-HE, an average simulated $\gamma_{LE,HE} = 0.015$ with a dispersion $\sigma_{LE,HE} = 0.076$; for ME-HE, an average simulated $\gamma_{ME,HE} = 0.022$ with a dispersion $\sigma_{ME,HE} = 0.081$. 
We conclude that the correlation between QPO intervals across non-overlapping energy bands found in our observational data is a real physical property, not an artifact of WWZ analysis.

\subsection{Spectral results}\label{subsec:sed}
For each observation, we built three sets of spectra (LE, ME and HE) of the QPO and non-QPO intervals (defined again from the WWZ maps). We then fitted them simultaneously for the three instruments, over the 2 -- 150 keV energy band (2 -- 10 keV for LE, 10 -- 30 keV for ME, and 30 -- 150 keV for HE), with standard models suitable to BH X-ray binaries. As an example, we illustrate here the results from ObsID P011466101502, chosen again because of its high signal-to-noise ratio; the spectral modelling results from the other observations are qualitatively similar.  

\begin{figure}
\centering
\includegraphics[scale=0.42]{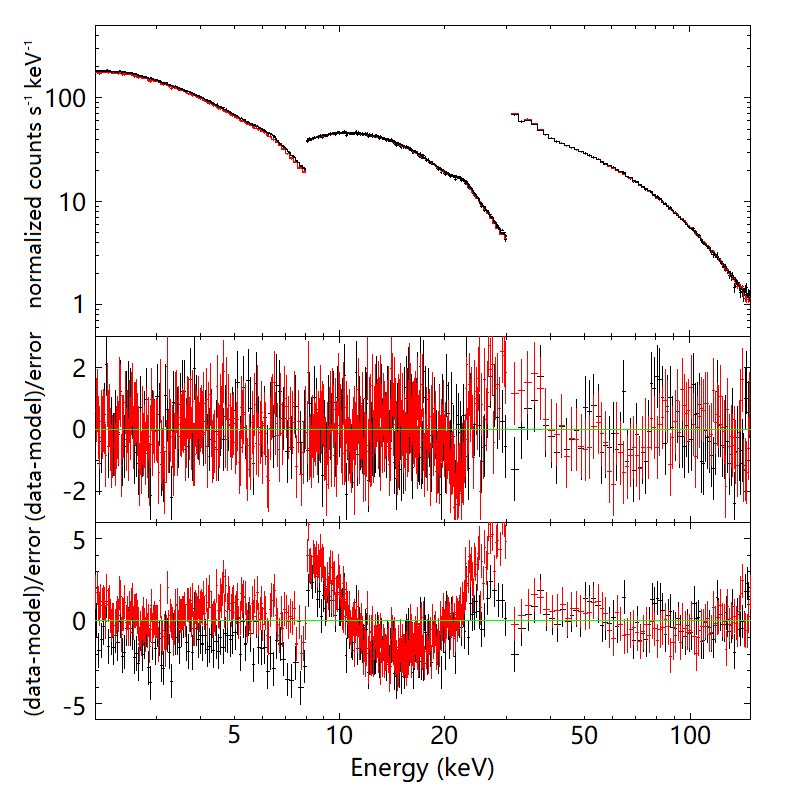}
\vspace{-0.4cm}
\caption{
{\it Insight-HXMT} spectrum from 2018 April 8 (ObsID P011466101502).
Top panel: the spectrum fitted with the model \emph{TBabs} $\times$ \emph{constant} $\times$ \emph{TBabs} $\times$ (\emph{diskbb} + \emph{gaussian} + \emph{relxill}); black datapoints are for observation intervals in which QPOs are detected, red datapoints for non-QPO intervals. Middle panel: residuals for the same spectral model. Bottom panel: residuals when the same spectrum is fitted with the simpler model \emph{constant} $\times$ \emph{TBabs} $\times$ ({\it diskbb} + {\it gaussian} + {\it cutoffpl}), that is without disk reflection. 
}
\label{energyspectrum}
\end{figure}

As a first trial, we used an absorbed disk-blackbody plus Comptonization plus Fe line model, {\it{const}} $\times$ {\it{TBabs}} $\times$ ({\it{diskbb}} $+$ {\it{gaussian}} $+$ {\it{cutoffpl}}), where the constant factor accounts for the systematic uncertainty in the normalization of the three instruments. The neutral absorption component {\it tbabs} \citep{Wilms2000} was fixed at $N_{\rm H} = 1.5 \times 10^{21}$ cm$^{-2}$ \citep{Uttley2018,You2021}. This model does not give an acceptable fit, with a reduced $\chi^2_{\nu} \gg 1$ (bottom panel of Fig.~\ref{energyspectrum}). The systematic residuals clearly suggest the presence of a reflection component with a characteristic bump around 30 -- 60 keV.

To account for the apparent reflection, we replaced the phenomenological power-law component with the more physical {\it relxill}\footnote{http://www.sternwarte.uni-erlangen.de/$\sim$ dauser/research/relxill/} model  \citep{Dauser2014, Garcia2014} version 1.2.0 \citep{Dauser2016}. \emph{Relxill} computes the combined direct and reflected spectrum of an incident primary continuum (a power-law with exponential high-energy cut-off) reflecting off an accretion disk. In our fits, we left the incident power-law photon index $\Gamma$ and cut-off energy $E_{\rm{cut}}$ as free parameters, as well as the Fe abundance $A_{\rm{Fe}}$, the maximum ionization $\xi$ of the disk, and the reflection fraction $R_{\rm f}$.
We fixed the power-law index of the disk emissivity profile at $q = 3$ for both the inner and outer disk regions, to reduce the number of free parameters. We also fixed the viewing angle at $i = 63^{\circ}$, based on the estimate of \cite{You2021}; in any case, the choice of viewing angle has only a small effect on the shape of the reflected spectrum. We also fixed the spin of BH at $a = 0.13$, based on the estimate of \citep{Guan2021}. We froze the inner disk radius parameter $R_{\rm in}$ at the innermost stable circular orbit (default in {\it{relxill}}); we also tested the possibility of leaving it as a free parameter, but we noticed that it would tend to the smallest possible value during the fitting process. The outer disk radius parameter $R_{\rm out}$ was fixed at a large value, outside the X-ray emitting region. Finally, we included a Gaussian line to model the Fe-K line emission around 6.4 keV, and a {\it{diskbb}} component to model the direct thermal emission from the disk. In summary, our final {\sc xspec} model was {\it{const}} $\times$ {\it{tbabs}} $\times$ ({\it{diskbb}} $+$ {\it{gaussian}} $+$ {\it{relxill}}). The best-fitting parameter values (Table \ref{tablelist}) were obtained with a Markov Chain Monte-Carlo algorithm. The model provides a good fit (Table \ref{tablelist} and top two panels of Fig.~\ref{energyspectrum}), with reduced $\chi^{2} \le 1.0$.

We stress that we are not trying to give a physical interpretation of the inflow structure or derive the system parameters of MAXI J1820$+$070 from our spectral modelling here. Our only motivation for this modelling is to test whether there is a significant spectral difference between sub-intervals with and without a QPO. Thus, we only aimed at obtaining a phenomenologically good fit over the {\it Insight-HXMT} energy range. For a more physical modelling of the spectral components, it is necessary to use also spectral data in the softer X-ray band, {\it e.g.}, from {\it XMM-Newton}, {\it Swift}, {\it NICER}. Such detailed analysis was carried out elsewhere \citep[{\it e.g.},][]{Shidatsu2019,Bharali2019,Kajava2019,Chakraborty2020,Paice2021,Dzielak2021,DeMarco2021,Axelsson2021,Prabhakar2022,Kawamura2022,Kawamura2023}. For example, we are aware that the inner radius of the geometrically thin accretion disk in the 2018 LHS, even near peak luminosity, was a few times larger than the innermost stable circular orbit $R_{\rm ISCO}$ \citep{Bharali2019,Shidatsu2019,DeMarco2021,Kawamura2023}. $R_{\rm ISCO}$ itself was estimated to be $\approx$ 70 km (\citealt{Shidatsu2019,Fabian2020}, using the system parameters of \citealt{Atri2020}). We are also aware that more than one Comptonization region ({\it e.g.}, a two-temperature corona, or a corona plus jet) may be required to fit the broadband X-ray spectrum \citep[{\it e.g.},][]{Chakraborty2020,Dzielak2021,Paice2021}.

Keeping in mind the previous caveats, our spectral results from an individual observation show that there is no substantial spectral difference between the times with and without a QPO. The only slight changes are in the photon index $\Gamma$, slightly steeper during QPO intervals (Table \ref{tablelist}), and (as a consequence of that) in the 30 -- 150 keV flux ($f_{30-150} = (2.7 \pm 0.1) \times 10^{-8}$ erg cm$^{-2}$ s$^{-1}$ in QPO intervals, and $f_{30-150} = (2.9 \pm 0.1) \times 10^{-8}$ erg cm$^{-2}$ s$^{-1}$ in non-QPO intervals). There is no difference (within the 90\% confidence limit) in the normalization of the reflection component, in the reflection fraction, and in the broadband luminosity. However, the error range of the best-fitting values for an individual observation are too large to detect subtle effects; for this aim, more work over the whole {\it Insight-HXMT} dataset is currently in preparation. 

\section{Discussion and conclusions}
\label{sec:summary}
We applied WWZ analysis to the {\it Insight-HXMT} light curve of MAXI J1820$+$070 in the LHS, during the 2018 March -- June outburst, to study the stability of flux oscillations in time--frequency space. We found that Type-C QPOs in this system are an intermittent rather than a persistent phenomenon. We showed that the average duration of each oscillatory interval is $\approx$ 5 QPO cycles, with a 3-$\sigma$ upper limit of $\approx$ 20 cycles. 

\begin{table*}
 \center
\caption{Best-fitting spectral parameters derived from our modelling of Obs. ID: P011466101502, for the full time interval and for those with and without a QPO signal. Errors are 90\% confidence limits for one interesting parameter. For the {\it diskbb} model, $R_{\rm in} \equiv 1.19 r_{\rm in} = \left(N_{\rm diskbb}\right)^{1/2}\,\left(\cos i \right)^{-1/2}\,\left(d/10 {\rm {kpc}}\right)$ \citep{Kubota1998}, and we have assumed a distance $d = 2.96$ kpc and a viewing angle $i = 63^{\circ}$.
$f_{2-150}$ is the observed flux in 2 -- 150 keV band, $L_{2-150}$ is the unabsorbed luminosity, for isotropic emission. Fluxes and luminosities were computed with the {\it cflux} model in {\sc xspec}. }
\begin{tabular}{lllllllll}
\hline
 Component & Parameter  & \multicolumn{3}{c}{Time Intervals} \\[4pt] 
 &   & Total & QPO Intervals & non-QPO Intervals \\ 
\hline
{\it constant} & $C_{LE}$ & $0.97_{-0.01}^{+0.01}$ & $1.10_{-0.01}^{+0.01}$ & $1.09_{-0.01}^{+0.01}$\\[4pt]
& $C_{ME}$ & $0.99_{-0.01}^{+0.01}$ & $1.03_{-0.01}^{+0.01}$ & $1.05_{-0.02}^{+0.01}$ \\[4pt]
  & $C_{HE}$ & [1.00] & [1.00] & [1.00]\\
\hline
{\it diskbb} & T$_{\rm in}$ (keV) & $0.65_{-0.02}^{+0.01}$ & $0.67_{-0.03}^{+0.01}$ & $0.61_{-0.03}^{+0.02}$ \\[4pt]
 & $N_{\rm diskbb}$ (km$^2$) & $3984_{-451}^{+598}$ &  $3816_{-584}^{+923}$  & $3423_{-269}^{+370}$\\[4pt]
 & $R_{\rm in}$ (km) &  $32.41_{-1.89}^{+2.35}$ & $31.72_{-1.78}^{+4.25}$ & $30.04_{-1.20}^{+1.58}$\\
\hline
{\it gaussian} & E$_{\rm Line}$ (keV) & $6.47_{-0.12}^{+0.18}$ & $6.54_{-0.34}^{+0.14}$  & $6.46_{-0.05}^{+0.07}$ \\[4pt]
& $\sigma$ (keV)& $0.34_{-0.04}^{+0.03}$ &  $0.31_{ -0.06}^{+0.02}$ &$0.21_{-0.07}^{+0.07}$\\[4pt]
 & $N_{\rm {line}}$ & $0.026_{-0.001}^{+0.001}$ & $0.036_{-0.002}^{+0.002}$  & $0.006_{-0.001}^{+0.002}$\\
\hline
{\it relxill} 
 & $\Gamma$ & $1.29_{-0.01}^{+0.01}$ &  $1.36_{-0.01}^{+0.01}$  & $1.23_{-0.01}^{+0.03}$\\[4pt]
 & log $\xi$ & $3.99_{-0.01}^{+0.02}$ & $3.97_{-0.02}^{+0.02}$ & $4.17_{-0.04}^{+0.04}$\\[4pt]
 & $Z_{\rm Fe}/Z_{\rm {Fe,}\odot}$ & $2.66_{-0.15}^{+0.20}$ & $2.68_{-0.22}^{+0.43}$ & $2.23_{-0.22}^{+0.43}$ \\[4pt]
 & $E_{\rm cut}$ (keV) & $188_{-40}^{+39}$ &  $201_{-39}^{+24}$  &  $144_{-11}^{+33}$\\[4pt]
 & refl\_frac & $0.42_{-0.06}^{+0.05}$ & $0.46_{-0.05}^{+0.06}$  & $0.38_{-0.06}^{+0.14}$ \\[4pt]
 & $N_{\rm {relxill}}$ & $0.075_{-0.002}^{+0.003}$ & $0.075_{-0.003}^{+0.002}$ &  $0.071_{-0.011}^{+0.006}$ \\
\hline
 & $\chi^{2}$/$\nu$ & 1352/1369 (0.99) & 1273/1369 (0.93) & 1218/1369 (0.89)\\
\hline
 & $f_{2-150}$ ($10^{-7}$ erg cm$^{-2}$ s$^{-1}$) & $1.07\pm 0.02$ & $1.07\pm 0.02$ & $1.07\pm 0.02$ \\[4pt]
 & $L_{2-150}$ ($10^{38}$ erg s$^{-1}$) & $1.1\pm 0.1$ & $1.1\pm 0.1$ & $1.1\pm 0.1$\\
\hline 
\end{tabular}
\label{tablelist} 
\end{table*}

Short-duration (intra-observation) QPO intervals were found in other Galactic BH X-ray binaries: in Cyg X-1 during a failed outburst \citep{Lachowicz2005}, and in XTE J1550$-$564 in a very high state \citep{Su2015}. In another Galactic BH in the very high state, GRS 1915$+$105, \cite{vandenEijnden2016} identified periods of high-amplitude coherent oscillations separated by time intervals in which the QPO loses phase coherence; this behaviour is probably analogous to the intermittent QPO behaviour highlighted here. An ultraluminous X-ray source in M\,51 (possibly a neutron star) exhibits intermittent oscillations with characteristic periods of $\sim$ 500 -- 700 s and a life time of a few cycles \citep{Urquhart2022}.   Intermittent QPOs with timescales of a few hours have been seen in Narrow-Line Seyfert I galaxies such as RE J1034$+$396 \citep{Middleton2011}, 1H 0707$-$495 \citep{Pan2016,Zhang2018} and Mrk 766 \citep{Zhang2017}. The sources mentioned above were of course in different accretion states ({\it e.g.}, near-Eddington for Narrow-Line Seyfert I galaxies, and super-Eddington for the ultraluminous source in M\,51), and their oscillations are likely to have different physical origin. We simply mention those other examples to stress that the intermittency characteristics of QPOs at different scales, and what distinguishes periods with and without a coherent oscillation, are still poorly known.

In the case of the intermittent QPOs of MAXI J1820$+$070 studied here, their frequency and the outburst state in which they were found leads to their classification as Type-C low-frequency QPOs. The X-ray photons in the oscillating component come from the Comptonizing region \citep{Chakrabarti2000}, either directly or via disk reflection, but the physical origin of the oscillation is still unclear \citep{Ingram2019}. One model \citep{Ingram2009} attributes them to the Lense-Thirring precession \citep{Lense1918} of the inner, geometrically thick, hot flow (corona or advective flow). Such precession causes periodic modulations of the projected area and maximum optical depth of the hot emission region with respect to our line of sight; it also modulates the flux of disk photons (seed thermal component) irradiating the Comptonizing region. The rigid precession model provides a physical explanation for intermittent QPOs, predicted to occur with a coherence Q approximately a few when the alignment time-scale $t_{\rm al}$ is only a few times the precession time-scale $t_{\rm pr}$. In this framework, the most important factors that determine the ratio $t_{\rm al}/t_{\rm pr} (\approx Q)$ are the outer radius of the geometrically thick part of the disk (the precessing region) and the BH spin \citep{Motta2018}. For a given spin parameter, larger radii of the thick precessing region correspond to a lower ratio of $t_{\rm al}/t_{\rm pr}$, and therefore to more intermittent, shorter-lived QPOs, down to the point where QPOs can no longer be produced ($t_{\rm al}/t_{\rm pr} < 2$). This also implies a minimum QPO frequency (maximum precessing radius) for any given spin. I particular, only slowly spinning BHs can produce Type-C QPOs with frequencies $\lesssim$0.1 Hz \citep[][, their Fig. 8]{Motta2018} as we observed in MAXI J1820$+$070. This is consistent with the slow spin proposed for MAXI J1820$+$070 from the continuum-fitting method \citep{Guan2021}. Alternatively, the precessing region might be temporarily blocked from our view by intervening disk material. However, the latter scenario is not consistent with our spectral results (Section 3.2), which show no change of the energy spectrum between QPO and non-QPO epochs.

Several other low-frequency QPO models are based on disco-seismic oscillations ({\it e.g.}, corrugation modes, inertial-acoustic modes, inertial-gravity modes); these are standing ways inside the disk, trapped between two characteristic radii, which give rise to characteristic resonant frequencies \citep[{\it e.g.},][]{Wagoner1999,Kato2001,Tsang2013,Ingram2019}. The advantage of such models is that they include mechanisms for excitation and damping of the various oscillatory modes \citep[{\it e.g.},][]{Li2003,Zhang2006,Tsang2009,Fu2009}. A comparison of the observed QPO life times with the predicted excitation and damping timescales is beyond the scope of this work. Similarly, we will leave to further work any discussion of excitation and damping in other well-known QPO models, such as those based on propagating oscillatory shocks near a centrifugal boundary layer \citep{Chakrabarti2008,Debnath2010}, or on the accretion-ejection instability in disks threaded by a strong poloidal magnetic field \citep{Tagger1999,Rodriguez2002,Tagger2007,Varniere2012}, or on disk oscillations and jet wobbling in response to instabilities near a magnetic recollimation zone \citep{Ferreira2022}.

Based on the {\it Insight-HXMT} detection of low-frequency QPOs in MAXI J1820$+$070 at energies as high as $\sim$ 200 -- 250 keV, and on the large soft lag (increasing with photon energy), \cite{Ma2021} argued that the QPOs in the LHS of this system are caused by Lense-Thirring precession of a compact jet. Furthermore, based on the evolution of the reflection fraction, \cite{You2021} suggested that the Comptonizing region in MAXI J1820$+$070 is a standing shock at the base of the jet, rather than a static hot region above the disk. Thus, we need to examine whether the jet precession scenario is consistent with the intermittent QPO properties seen in the WWZ maps. We already noted that the main difference between the two regimes is a reduced rms amplitude of the oscillation. In the precessing jet model \cite{Ma2021}, the observed fractional rms amplitude of the low-frequency QPOs is a function of jet speed (via Doppler beaming of the X-ray photons): smaller jet speeds lead to fainter QPOs. Therefore, we speculate that the discontinuity of the QPO signal may be caused by changes in the jet velocity on timescales of a few tens to a few hundred seconds. For example, the average fractional rms of $\approx$ 11.3\% measured for QPO intervals (Section 3.1) corresponds to a jet speed of $\approx$ 0.52$c$ (Extended Data Fig.~8 in \citealt{Ma2021}); for the non-QPO intervals, the observed rms upper limit corresponds to a 3$\sigma$ upper limit of $\approx$ 0.47$c$.

The characteristic timescale of jet speed variability ($\sim$ 5 -- 20 QPO cycles) may, in turn, be determined by mass loading, hence by variations of the accretion rate, or other inflow properties such as inner-disk or boundary layer oscillations. A discussion of this disk/jet coupling is beyond the scope of this preliminary work. As an empirical comparison, the jet speed in the best studied Galactic microquasar, SS\,433, was observed to vary between $\approx$ 0.21$c$ and $\approx$ 0.32$c$, correlated with the collimation angle \citep{Blundell2005,Blundell2007,Jeffrey2016}.

Recently, \cite{Gao2023} carried out a timing study of Type-C QPOs in the 2018 March -- June outburst of MAXI J1820$+$070, also based on {\it Insight-HXMT} data. Their methodology is somewhat complementary to our analysis: they selected sub-intervals with strong QPOs and extracted phase-resolved spectra around the peaks and the troughs of the oscillation. Their main finding is that QPOs are dominated by an oscillation of the reflection fraction, for photons with energies $\la$30 keV, and by an oscillation of the direct Comptonized component above those energies. We examined whether our proposed scenario of variable speed in a precessing jet is consistent with those results. At first sight (as argued by \citealt{Gao2023}), a precessing jet alone seems to be unable to create an oscillating reflection fraction, because the solid angle of the disk seen by comptonized photons does not change during the jet precession. However, a misaligned jet will always illuminate more strongly one side of the disk. During the precession cycle, Doppler boosting in the disk will increase the observed reflection component when the jet is illuminating the approaching side of the rotating disk, and decrease the reflection component when it illuminates the receding side \citep{Ingram2017}. Thus, we conclude that a Lense-Thirring precession (at constant jet speed) is perfectly consistent with the observed X-ray properties of the oscillation during QPO sub-intervals; a change in jet speed is consistent with the observed strengthening and weakening (or disappearance) of the QPO over characteristic timescales of tens to hundreds of seconds, highlighted in this study. 

\section*{Acknowledgements}
This work is supported by the National Key R\&D Program of China (2021YFA0718500) and the National Natural Science Foundation of China under grants 12203029, U1838201, U1838202, U1938101, 12073029, 11733009, 12233002 and U2031205.
This work made use of data from the \emph{Insight}-HXMT mission, a project funded by China National Space Administration (CNSA) and the Chinese Academy of Sciences (CAS). We thank Sandip Chakrabarti for discussions on QPO models.

\bibliographystyle{aa}
\bibliography{myBiblio}

\label{lastpage}
\end{document}